# Is Einstein's relation D = kTU always true?

Roumen Tsekov
Department of Physical Chemistry, University of Sofia, 1164 Sofia, Bulgaria

The validity of Einstein's fluctuation-dissipation relation is discussed in respect to the type of relaxation in an isothermal system. The first model, presuming isothermic fluctuations, leads to the Einstein formula. The second model supposes adiabatic fluctuations and yields another relation between the diffusion coefficient and mobility of a Brownian particle. A new approach to relaxations in quantum systems is also proposed that demonstrates applicability only of the adiabatic model for description of the quantum Brownian dynamics.

Ninety years ago, Einstein [1] noted that the diffusion coefficient $D$ of a Brownian particle moving in a fluid is related to its mobility $U$ via the relation

$$D = kTU \qquad (1)$$

where $k$ is the Boltzmann constant and $T$ is the temperature of the fluid. The scope of the present paper is to show that the Einstein relation (1) is not general and there are certain conditions under which it is not valid. To make the consideration plausible, let us start with the derivation of the Einstein formula once again. Imagine a rested fluid in contact to a thermostat with constant temperature. The observation follows the motion of a particle with mass $m$ immersed in the fluid. If this motion is much slower than temperature relaxations in the fluid (this occurs when the Brownian particle is much heavier than fluid particles), one can assume that fluctuations around the Brownian particle are isothermic. Hence, the dynamics of the Brownian particle probability density will obey the following equation (see Appendix)

$$\partial_t \rho = \partial_r \cdot [\rho \partial_r (V + \mu)/b_T] \qquad (2)$$

where $\rho$ is the probability density to find the Brownian particle in a given point $r$ at the moment $t$, $b_T$ is the friction constant of isothermic relaxation, $V$ is an external potential and $\mu$ is the chemical potential or molar Gibbs energy. The fact that the chemical potential gradient is the average representative of isothermic fluctuation forces is clearly expressed in the thermodynamic Gibbs-Duhem equation $\partial_r \cdot \Pi = \rho \partial_r \mu$ at constant $T$.

Since the Brownian particle is only one, the chemical potential should be that for a dilute solution

$$\mu = \mu_0 + kT \ln \rho \tag{3}$$

and thus one obtains from eqs. (2) and (3) the well-known diffusion equation

$$\partial_t \rho = \partial_r \cdot (\rho \partial_r V + kT \partial_r \rho) / b_T \tag{4}$$

It is a particular example of Fokker-Planck equations named after Smoluchowski [2]. As expected, eq. (4) provides as its equilibrium solution the Boltzmann distribution

$$\rho_{eq} \sim \exp(-\beta V) \tag{5}$$

where $\beta \equiv 1/kT$. In the case of a constant external force $F = -\partial_r V$, the solution of eq. (4) is a normal distribution density with mean value and dispersion given by

$$<r> = Ft/b_T \qquad \sigma^2 \equiv <(r-<r>)^2>/3 = 2D_T t \tag{6}$$

where the diffusion coefficient is defined as

$$D_T \equiv kT/b_T \tag{7}$$

Introducing the mobility $U$ via the relation $<r> = UFt$, it follows from eq. (6) that $U = 1/b_T$. Hence, eq. (7) coincides with the Einstein formula (1).

### A model violating the Einstein relation

The above derivation of Einstein's relation seems to be quite general and it is not obvious to see limitations of the model. However, there is a case when the presented theory failures. Let us imagine a Brownian particle having mass many times smaller than that of fluid particles. Now, one can expect that fluid relaxations cannot follow the quick perturbations caused by the particle and the local temperature can largely fluctuate. This means that even if the whole system is isothermal, the fluctuations around the Brownian particle are adiabatic. In this case, the probabilistic equation corresponding to eq. (2) gets the form (see Appendix)

$$\partial_t \rho = \partial_r \cdot [\rho \partial_r (V + h)/b_A] \tag{8}$$

where $b_A$ is the adiabatic friction constant and the molar enthalpy $h$ is the average representative of the adiabatic fluctuation forces since $\partial_r \cdot \Pi = \rho \partial_r h$ at constant entropy. The main differ-ence between eq. (2) and eq. (8) is that the chemical potential $\mu$ is replaced by the molar enthalpy $h$. However, there is a thermodynamic relationship between these two quantities

$$h = \partial_\beta (\beta \mu) \tag{9}$$

known as the Gibbs-Helmholtz equation. Using eqs. (3), (8) and (9), one can obtain a new equation for the adiabatic evolution of the probability density

$$\partial_t \rho = \partial_r \cdot [\rho \partial_r (V + \partial_\beta \ln \rho)/b_A] \tag{10}$$

This equation expresses the two main characteristics of the process, i.e. adiabatic relaxation and isothermal equilibrium. For this reason, the equilibrium distribution provided by eq. (10) is Boltzmann's one (5) again.

If the external force is constant, one should look for the solution of eq. (10) in the form of a normal distribution density

$$\rho = \exp[-(r - <r>)^2 / 2\sigma^2] / \sqrt{(2\pi\sigma^2)^3}$$

Introducing it in eq. (10), one obtains that the mean value $<r>$ and dispersion $\sigma^2$ obey the relations

$$b_A <r> = Ft - \partial_\beta <r> / 2D_A \qquad \sigma^2 = 2D_A t \tag{11}$$

where the adiabatic diffusion coefficient $D_A$ is related to the friction constant via the expression

$$\partial_\beta (1/D_A) = b_A \tag{12}$$

The last equation demonstrates an additional property of $D_A$ compared to $D_T$. Both diffusion coefficients are positive but because $b_A > 0$ it follows from eq. (12) that $D_A$ is an increasing function of the temperature. Such a restriction for $D_T$ does not exist.

If one introduces the particle mobility $U$ via $<r> = UFt$, eq. (11) can be transformed by using eq. (12) to

$$U \partial_\beta (1/D_A) = 1 - \partial_\beta U / 2D_A$$

This differential equation can be solved under the obvious condition $D_A(\beta \to 0) \to \infty$ following from the monotonously increasing dependence of $D_A$ on $T$ noted above. The solution is

$$D_A = \sqrt{U} / \int_0^\beta d\beta / \sqrt{U} \tag{13}$$

It is easy to obtain by combining eqs. (12) and (13) how the adiabatic friction constant can be calculated from the experimentally measurable mobility.

As is seen from eq. (13), the relation between $D_A$ and $U$ differs from the Einstein formula (1). The reason is the adiabatic character of the fluctuations around the Brownian particle. There is a question remaining unopened: is there a link between the mobility in the two cases considered above? Due to usually small mean velocity of the Brownian particle as compared to the thermal velocity of the fluid particles, one could suppose that the mobility coincides in both cases. Hence, since $b_T$ is a well-known quantity (for instance the Stokes formula), this assumption will help to obtain expressions for unknown $D_A$ and $b_A$. However, such an argument is correct if the temperature fluctuations affect the Brownian motion in a linear way that their average contribution will cancel. Since this is not a general property of the system, the coincidence of mobility is not obvious.

Finally, the fluctuations around the Brownian particle can be neither isothermic nor adiabatic. In this case a natural extension of eqs. (4) and (10) is

$$\partial_t \rho = \partial_r \cdot \{\rho \partial_r [V + \alpha kT \ln \rho + (1-\alpha)\partial_\beta \ln \rho]/b\} \tag{14}$$

where the parameter $\alpha$ could depend on temperature. The parameters in eq. (14) are chosen in such a way that the equilibrium solution is the Boltzmann distribution (5). In the case of a constant external force, one can obtain from eq. (14) the following two relations

$$b = \alpha kT/D + (1-\alpha)\partial_\beta(1/D) \qquad\qquad bU = 1 - (1-\alpha)\partial_\beta U / 2D$$

which contain the two particular cases considered before: isothermic ($\alpha = 1$) and adiabatic ($\alpha = 0$) relaxations of the Brownian particle. One could also generalize the present point by application of the idea to the evolution of probability density $f(p,r,t)$ in the Brownian particle phase space. The corresponding equation should be

$$\partial_t f + p \cdot \partial_r f/m - \partial_r V \cdot \partial_p f = b \partial_p \cdot [pf/m + \alpha kT \partial_p f + (1-\alpha) f \partial_p \partial_\beta \ln f]$$

and its equilibrium solution is the Maxwell-Boltzmann distribution. In the isothermic case ($\alpha = 1$) this equation coincides with the Klein-Kramers equation [2, 3].

## Application to quantum Brownian motion

As was demonstrated, deviations from the Einstein formula (1) could take place if the particle is sufficiently light. A good example for the adiabatic model should be then the diffusion of electrons in liquids. In this case, however, the quantum effects are important and have to be accounted for. We shall follow an approach to the problem developed by us before [4]. In the framework of the quantum mechanics, the motion of a single particle in an external potential field $V$ is described by the Schrödinger equation

$$i\hbar \partial_t \psi = \hat{H}\psi = -\hbar^2 \partial_r^2 \psi / 2m + V\psi \qquad (15)$$

The wave function $\psi$ can be presented in a polar form

$$\psi = \sqrt{\rho}\exp(i\Omega)$$

where $\rho = |\psi|^2$ is probability density to find the particle at a given point at time $t$. Substituting this presentation into eq. (15), the latter splits into the following two equations

$$\partial_t \rho + \partial_r \cdot (\rho v) = 0 \qquad (16)$$

$$m\partial_t v + mv \cdot \partial_r v = -\partial_r (V + Q) \qquad (17)$$

corresponding to real and imaginary part. The velocity in the probability space $v$ is introduced here via the relation $v = \hbar \partial_r \Omega / m$. Equation (16) is the continuity equation, while eq. (17) represents the momentum balance of the particle. The last term $Q \equiv -\hbar^2 \partial_r^2 \sqrt{\rho}/2m\sqrt{\rho}$ in the brackets is known as the quantum potential and it disappears in the classical limit $\hbar \to 0$.

The two equations above describe motion of a particle in vacuum. In the case of Brownian motion in a fluid, two new forces should be added to eq. (17): a friction force, proportional to the velocity $v$, and a driving force, generated by fluctuations in the surrounding medium. Hence, in a common case, a natural extension of eq. (17) reads

$$m\partial_t v + mv \cdot \partial_r v + bv = -\partial_r [V + Q + \alpha kT \ln \rho + (1-\alpha)\partial_\beta \ln \rho]$$

Now, neglecting the first two inertial terms in respect to the friction, one obtains the following expression for the velocity

$$v = -\partial_r[V + Q + \alpha kT \ln\rho + (1-\alpha)\partial_\beta \ln\rho]/b$$

which, introduced in eq. (16), leads to

$$\partial_t\rho = \partial_r \cdot \{\rho\partial_r[V + Q + \alpha kT \ln\rho + (1-\alpha)\partial_\beta \ln\rho]/b\} \tag{18}$$

Equation (18) is the quantum analogue of eq. (14) and reduces to the latter in the classical limit $\hbar \to 0$. According to quantum statistical thermodynamics, the equilibrium solution of eq. (18) should be the canonical Gibbs distribution

$$\rho_n(r) = \exp(-\beta E_n)\varphi_n^2(r) / \sum_n \exp(-\beta E_n) \tag{19}$$

where $\{E_n, \varphi_n\}$ is the set of eigenvalues and eigenfunctions of the Hamiltonian $\hat{H}$. The distribution (19) is solution of eq. (18) only if $\alpha = 0$. This fact shows that for quantum particles the adiabatic model is not only relevant but it is the only possible one. This is not surprising since quantum particles possess very small masses and the surrounding fluid could not follow their quick motion as was mentioned before. A general equilibrium solution of eq. (18) at $\alpha = 0$ is

$$\rho_{eq}^{1/2} \sim \exp(-\beta\hat{H}/2) \tag{20}$$

which reduces to a constant at $V = 0$ and recovers the Boltzmann distribution (5) in the classical limit $\hbar \to 0$.

**Appendix**

Let $R(t)$ is a random trajectory of the Brownian particle. Then, the corresponding probability density is given by

$$\rho(r,t) = <\delta(r-R)> \tag{A1}$$

where the brackets $<\cdot>$ denote statistical average over all possible realizations of $R$ and $\delta(\cdot)$ is the Dirac delta function. Taking a derivative of eq. (A1) in respect to time yields the well-known continuity equation

$$\partial_t \rho = -\partial_r \cdot <\dot{R}\delta(r-R)> \equiv -\partial_r \cdot (\rho v) \qquad (A2)$$

In this manner the macroscopic momentum balance could be also derived

$$\partial_t(\rho v) = <\ddot{R}\delta(r-R)> - \partial_r \cdot <\dot{R}\dot{R}\delta(r-R)> \qquad (A3)$$

Let the random process $R$ obeys the Langevin equation [5]

$$m\ddot{R} + b\dot{R} = -\partial_R V + f$$

where $b$ is the friction constant, $V$ is an external potential and $f$ is the stochastic force. Introducing the particle acceleration from here in eq. (A3), the latter changes to

$$m\partial_t(\rho v) = -\rho\partial_r V - b\rho v - \partial_r \cdot <m\dot{R}\dot{R}\delta(r-R)> + <f\delta(r-R)> \qquad (A4)$$

It can be shown that the last term here is zero [3]. On the other hand, the derivative on the left hand side can be expressed using eq. (A2) in the form

$$\partial_t(\rho v) = \rho\partial_t v + v\partial_t \rho = \rho\partial_t v + \rho v \cdot \partial_r v - \partial_r \cdot (\rho vv) \qquad (A5)$$

Hence, the combination of eqs. (A4) and (A5) provides

$$m\partial_t v + mv \cdot \partial_r v + bv = -\partial_r V - \partial_r \cdot \Pi/\rho \qquad (A6)$$

where $\Pi$ is defined via the relation [3]

$$\Pi \equiv <m(\dot{R}-v)(\dot{R}-v)\delta(r-R)>$$

One can easily recognize in $\Pi$ the osmotic pressure tensor due to the particle.
In the case of high friction, the inertial terms is eq. (A6) can be neglected as compared to the friction force and the velocity $v$ acquires the simple form

$$v = -(\partial_r V + \partial_r \cdot \Pi/\rho)/b$$

Substituting this expression in eq. (A2) yields

$$\partial_t \rho = \partial_r \cdot (\rho \partial_r V + \partial_r \cdot \Pi)/b$$

For the further analysis additional specification of the dependence of the osmotic pressure on density $\rho$ is required.